\documentclass[12pt]{cernart}
\usepackage{epsfig}
\newcommand{\e}{\mathrm{e}}
\newcommand{\Nb}{\mbox{$\overline{N}$}}
\newcommand{\be}{\begin{equation}}
\newcommand{\ee}{\end{equation}}
\newcommand{\Pkab}{\mbox{$\mathrm{\overline{K}{}^0}$}}
\newcommand{\Pkao}{\mbox{$\mathrm{K}{}^0$}}
\newcommand{\Pkal}{\mbox{$\mathrm{K_L}$}}
\newcommand{\Pkas}{\mbox{$\mathrm{K_S}$}}
\newcommand{\gs}{\Gamma_\mathrm{S}}
\newcommand{\gl}{\Gamma_\mathrm{L}}
\newcommand{\Mkao}{\mbox{$\mathrm{M_{K^0}}$}}
\newcommand{\deltam}{\Delta m }
\newcommand{\ms}{m_\mathrm{S}}
\newcommand{\ml}{m_\mathrm{L}}
\newcommand{\metapm}{|\eta_{+-}|}
\newcommand{\phipm}{\phi_{+-}}
\newcommand{\pppm}{\pi^+\pi^- } 
\newcommand{\Ppb}{$\overline{\mathrm{p}}$}
\def\lsim{\mathrel{\rlap {\raise.5ex\hbox{$ < $}}
    {\lower.5ex\hbox{$\sim$}}}}
\begin{document}
\begin{titlepage}
\docnum{CERN-EP/99--22}
\docnum{hep-ex/9903005}
\date{February 15, 1999}
\title{Tests of the Equivalence Principle with
       Neutral Kaons }
\vspace*{-0.5cm}
\collaboration{CPLEAR Collaboration}
\begin{Authlist}
A.~Apostolakis\Iref{a1},
E.~Aslanides\Iref{a11}, 
G.~Backenstoss\Iref{a2}, 
P.~Bargassa\Iref{a13},
O.~Behnke\Iref{a17}, 
A.~Benelli\Iref{a2}, 
V.~Bertin\Iref{a11}, 
F.~Blanc\IIref{a7}{a13},
P.~Bloch\Iref{a4}, 
P.~Carlson\Iref{a15}, 
M.~Carroll\Iref{a9},
E.~Cawley\Iref{a9},
G.~Chardin\Iref{a14} 
M.B.~Chertok\Iref{a3},
M.~Danielsson\Iref{a15}, 
M.~Dejardin\Iref{a14},
J.~Derre\Iref{a14}, 
A.~Ealet\Iref{a11}, 
C.~Eleftheriadis\Iref{a16}, 
L.~Faravel\Iref{a7},
W.~Fetscher\Iref{a17}, 
M.~Fidecaro\Iref{a4},
A.~Filip\v ci\v c\Iref{a10}, 
D.~Francis\Iref{a3}, 
J.~Fry\Iref{a9},
E.~Gabathuler\Iref{a9}, 
R.~Gamet\Iref{a9}, 
H.-J.~Gerber\Iref{a17},
A.~Go\Iref{a4}, 
A.~Haselden\Iref{a9},
P.J.~Hayman\Iref{a9}, 
F.~Henry-Couannier\Iref{a11},
R.W.~Hollander\Iref{a6}, 
K.~Jon-And\Iref{a15}, 
P.-R.~Kettle\Iref{a13},
P.~Kokkas\Iref{a4}, 
R.~Kreuger\Iref{a6},
R.~Le Gac\Iref{a11}, 
F.~Leimgruber\Iref{a2}, 
I.~Mandi\' c\Iref{a10}, 
N.~Manthos\Iref{a8},
G.~Marel\Iref{a14}, 
M.~Miku\v z\Iref{a10}, 
J.~Miller\Iref{a3}, 
F.~Montanet\Iref{a11},
A.~Muller\Iref{a14},
T.~Nakada\Iref{a13}, 
B.~Pagels\Iref{a17},
I.~Papadopoulos\Iref{a16},
P.~Pavlopoulos\Iref{a2},
G.~Polivka\Iref{a2}, 
R.~Rickenbach\Iref{a2}, 
B.L.~Roberts\Iref{a3},
T.~Ruf\Iref{a4}, 
L.~Sakeliou\Iref{a1}, 
M.~Sch\"afer\Iref{a17}, 
L.A.~Schaller\Iref{a7}, 
T.~Schietinger\Iref{a2},
A.~Schopper\Iref{a4}, 
L.~Tauscher\Iref{a2},
C.~Thibault\Iref{a12}, 
F.~Touchard\Iref{a11}, 
C.~Touramanis\Iref{a9},
C.W.E.~Van Eijk\Iref{a6},
S.~Vlachos\Iref{a2}, 
P.~Weber\Iref{a17}, 
O.~Wigger\Iref{a13}, 
M.~Wolter\Iref{a17},
D.~Zavrtanik\Iref{a10},
D.~Zimmerman\Iref{a3}\\[3mm]
and \\[3mm]
John~Ellis\Iref{a18},
N.E.~Mavromatos\Iref{a19},
D.V.~Nanopoulos\Iref{a20}\\[4mm]
\end{Authlist}
  \begin{abstract}
    We test the Principle of Equivalence for particles and
    antiparticles, using CPLEAR data on tagged \Pkao\ and \Pkab\ 
    decays into $\pi^+ \pi^-$. For the first time, we search for
    possible annual, monthly and diurnal modulations of the 
    observables $|\eta_{+-}|$ and $\phi _{+-}$, 
    that could be correlated with variations in
    astrophysical potentials. Within the accuracy of CPLEAR, 
    the measured values of $|\eta _{+-}|$ and $\phi _{+-}$ 
    are found not to be correlated with changes of the
    gravitational potential. We analyze data assuming effective 
    scalar, vector and tensor interactions, and we conclude that
    the Principle of Equivalence between particles and antiparticles 
    holds to a level of $6.5$, $4.3$ and $1.8 \times 10^{-9}$,
    respectively, for scalar, vector and tensor potentials
    originating from the Sun with a range much greater than the
    distance Earth-Sun.  We also study energy-dependent
    effects that might arise from vector or tensor interactions. 
    Finally, we compile upper limits on the
    gravitational coupling difference between \Pkao\ and \Pkab\ 
    as a function of the scalar, vector and tensor interaction
    range.
   
\end{abstract}
\submitted{(submitted to Phys. Lett. B)}
\Instfoot{a1}{University of Athens, Greece}
\Instfoot{a2}{University of Basle, Switzerland}
\Instfoot{a3}{Boston University, USA}
\Instfoot{a4}{CERN, Geneva, Switzerland}
\Instfoot{a5}{LIP and University of Coimbra, Portugal}
\Instfoot{a6}{Delft University of Technology, Netherlands}
\Instfoot{a7}{Uni\-ver\-sity of Fribourg, Switzerland}
\Instfoot{a8}{Uni\-ver\-sity of Ioannina, Greece}
\Instfoot{a9}{Uni\-ver\-sity of Liverpool, UK}
\Instfoot{a10}{J.~Stefan Inst.\ and Phys.\ Dep., University of Ljubljana,
 Slovenia}
\Instfoot{a11}{CPPM, IN2P3-CNRS et Universit\'e d'Aix-Marseille II, France}
\Instfoot{a12}{CSNSM, IN2P3-CNRS, Orsay, France}
\Instfoot{a13}{Paul Scherrer Institut (PSI), Villigen, Switzerland}
\Instfoot{a14}{CEA, DSM/DAPNIA, CE-Saclay, France}
\Instfoot{a15}{Royal Institute of Technology, Stockholm, Sweden}
\Instfoot{a16}{University of Thessaloniki, Greece}
\Instfoot{a17}{ETH-IPP Z\"urich, Switzerland}
\Instfoot{a18}{Theory Division, CERN, 1211 Geneva 23, Switzerland}
\Instfoot{a19}{Dept. of Physics (Theoretical Physics), University of Oxford, 
    1 Keble Road, Oxford OX1 3NP, U.K.}
\Instfoot{a20}{Academy of Athens, 10679 Athens, Greece, 
    Center for Theoretical Physics, Department of Physics, Texas
    A\&M University,
    College Station, TX 77843--4242, USA, and
    Astroparticle Physics Group,
    Houston Advanced Research Center, The Mitchell Campus, 
    The Woodlands, TX 77381, USA}
  \vspace{0.5cm}

\end{titlepage}
\vfill\eject
\setcounter{page}{1}
\pagestyle{plain}

\section{Introduction}

The neutral kaon system is a very sensitive laboratory 
for the exploration of possible differences between matter 
and antimatter. Indeed, it is the only system where any matter-antimatter 
difference has been seen, which is 
conventionally ascribed to a CP-violating term
in the \Pkao -- \Pkab\ mass matrix. 
Searches for other asymmetries between \Pkao\ and \Pkab\ 
have also been conducted, notably to set upper limits on effects 
violating CPT invariance \cite{cptkaon}. These have included a
possible \cite{cptconv} CPT-violating \Pkao -- \Pkab\ mass difference,
$\delta m$, and width difference, $\delta \Gamma $, \cite{pdg96, cplearcptqmm}
and stochastic CPT
violation of the form that may appear if neutral kaons 
should be described as an open 
quantum-mechanical system \cite{ehns,emn,cptcplear}. 
These searches have been motivated in part by suggestions 
that some form of CPT violation may occur 
in a quantum theory of gravity \cite{cptgrav}.

It has also been suggested that some quantum theory of gravity
might entail an apparent violation of the Principle of
Equivalence between matter and antimatter particles, 
due for instance to the possible exchange of a light vector particle with 
interactions of gravitational strength, a ``graviphoton'' \cite{graviphoton}. 
Such a deviation from the Principle of Equivalence would imply a 
different gravitational coupling between particle and antiparticle, 
independently of the universality of the coupling between matter
and gravity \cite{equiv}. 

The aspect of the Principle of Equivalence concerning the universality
of the coupling between the graviton and matter has been verified by
many and diverse experiments, which, over the last four decades,
have placed stringent limits on possible deviations from General
Relativity. These include phenomenological searches for
non-universality in free fall \cite{su94}, 
the fractional difference in the acceleration of the Earth and Moon in the 
Sun's gravitational field using lunar laser ranging \cite{will96}, 
the universality of the gravitational red shift using very stable clocks
on aircraft, rockets and satellites \cite{ves80},
the spatial anisotropy of nuclear energy levels \cite{chup89},
and upper limits on
time variations in the basic coupling constants of the Standard 
Model \cite{pres95}.

The other aspect of the Equivalence Principle, that of the 
universality  of the gravitational coupling between matter and antimatter,
can be tested by particle--antiparticle comparisons, such as 
\Pkao -- \Pkab\ ~\cite{cplearcptqmm} and p -- \Ppb\
~\cite{gab90} mass-difference measurements, which 
can be interpreted as tests of the Principle of Equivalence
under the assumption of exact CPT symmetry \cite{hug91}. 

Data from the CPLEAR experiment have been used previously to set the 
most precise available upper limits on a  possible CPT-violating
\Pkao -- \Pkab\ mass difference,
$\delta m$, and width difference, $\delta \Gamma$, \cite{cplearcptqmm} 
and on stochastic CPT-violating parameters \cite{cptcplear}. 
These data are used in the present paper as a function of time,
in order to provide tests  of the second aspect 
of the Principle of Equivalence comparing \Pkao\ and \Pkab . After 
presenting the formalism, we report on a systematic 
search for possible annual, monthly and diurnal 
modulations of the observables $|\eta_{+-}|$ and $\phi_{+-}$, as
functions of the known variations of the astrophysical potentials. 
The upper limits so obtained can be used to constrain differences 
between \Pkao\ and \Pkab\ interactions with a background 
field, as a function of its intrinsic spin and range.
We use these bounds to discuss the possibility that all of
the observed CP-violating effects could be due to an astrophysical
scalar field \cite{Bell}. In addition, the \Pkao -- \Pkab\
mass difference provided by the CPLEAR experiment, which is ten
orders of magnitude more precise than the available p -- \Ppb\ 
mass difference, puts stronger upper limits on the violation of the
Principle of Equivalence for longer-range interactions, 
by considering galactic and extragalactic background fields. 
Finally, we comment on the possible combination 
of CPLEAR data with higher-energy data to constrain more tightly
interactions with vector or tensor background fields. 

Although limits on differences between \Pkao\ and \Pkab\ interactions
with a background field,  
arising from limits on the \Pkao -- \Pkab\ mass
difference, have already been reported in the literature \cite{hughes},
limits derived from 
studies of the observables $|\eta_{+-}|$ and $\phi_{+-}$,
in relation with
the known time variations of the astrophysical potentials,
are, to our knowledge, entirely new. We note that limits derived
from possible modulations of $|\eta_{+-}|$ and $\phi_{+-}$ related to
the modulation of astrophysical potentials do not depend on
the less well known  galactic and extragalactic 
potentials, those being constant during the lifetime of the experiment. 

\section{Formalism for Equivalence Tests} 

The standard treatment of mass mixing in the neutral kaon system is based 
on the following parameterization
\be
{\cal M}=\left(\begin{array}{cc} \Mkao \qquad \deltam /2   \\
    \deltam /2 \qquad \Mkao \end{array}\right)
\label{dm}
\ee
where 
\Mkao\  is the kaon inertial mass, and $\deltam = \ml - \ms$
is the mass difference between the long- and short-lived
neutral kaons. Non-invariance under CPT would induce, in principle,
a mass  difference, $\delta m$, between the \Pkao\ and \Pkab,
which is limited to $\delta m \leq 3.5 \times 10^{-19}~{\rm GeV}$
($95\%$ CL) \cite{cplearcptqmm}.
We assume in (\ref{dm}) and the rest of this paper the equality
of the  inertial masses and widths of
\Pkao\ and \Pkab . Also, we work throughout in the conventional
quantum-mechanical formalism.

A violation of the Principle of Equivalence could arise from the 
possibility that the \Pkao\ and \Pkab\ might have
different interactions with the surrounding astrophysical matter
distribution via background fields of tensor, vector and scalar types. 
Possible sources of such interactions are astrophysical bodies
at generic distances $r$. One parametrizes the possible magnitudes
of their effects relative to the conventional gravitational potential
$U=G_NM/r$ due to a body of mass $M$, where $G_N$ is the Newton's
constant, by introducing relative couplings
$g$ and ${\overline g}$ for the \Pkao\ and \Pkab,
respectively. 
Any matter-antimatter difference in the couplings
of a field of spin $J$ and effective range $r_J$, ${(g - {\overline g})_J}$,
would entail an effective violation of the Principle of Equivalence,
since it would violate the universality of free fall for
the \Pkao\ and \Pkab. Such effects depend explicitly on the
potential $U$ of the gravitational interaction and have an exponential
dependence on the field effective range $r_J$. 

In the case of a tensor gravitational interaction
the effective \Pkao -- \Pkab\ mass difference
acquires the form \cite{hughes}:
\be
\delta m_{eff} = \Mkao {(g -{\overline g})}_2
\frac{U}{c^2} (1 + v^2/c^2)\gamma ^2 \e ^{-r/r_2} ~,
\label{diff}
\ee
where 
$\gamma = 1/\sqrt{1-v^2/c^2}$.
In the case of a vector interaction, we assume that any vector interaction
couples to the four-velocity field of the neutral kaon via
$\int dt' g_V V_\mu {\dot x}^\mu$, with $V_\mu=(U/c^2, 0, 0, 0)$
for static sources, leading to 
\be
\delta m_{eff} = \Mkao {(g -{\overline g})}_1
\frac{U}{c^2} \gamma \e ^{-r/r_1}~.
\label{diffvect}
\ee
Finally, in the case of
scalar interactions, we assume a ``dilaton-like'' coupling 
to the trace of the stress-tensor of the kaon system, as would
be the case for a Brans-Dicke scale factor, leading to
\be
\delta m_{eff} = \Mkao {(g -{\overline g})}_0
\frac{U}{c^2} \e ^{-r/r_0} ~.
\label{diffscal}
\ee
We emphasize the different functional forms for the tensor, 
vector and scalar interactions
given in Eqs. (\ref{diff}), (\ref{diffvect}) and (\ref{diffscal}),
in particular the energy independence of the scalar interaction.
Within the framework of a conventional quantum field
theory we would expect that ${(g - {\overline g})}_{J} = 0$ for $J=0$ or $2$,
whereas ${(g - {\overline g})}_1 \ne 0$ could be generated by
graviphoton exchange \cite{graviphoton}.

Such \Pkao -- \Pkab\ mass differences modify the values 
of the CP-violation observables:
\be 
|\eta_{+-}|^2 \simeq 
|\epsilon |^2 + \frac{(\delta m _{eff})^2}{8(\Delta m)^2}
\label{etapm}
\ee
and 
\be
\tan\phi _{+-} \simeq
\tan \phi _{sw} + \frac{1}{2\sqrt{2}} \frac{\delta m _{eff}}
{\Delta m |\epsilon|} (\tan^2 \phi _{sw} + 1) ~, 
\label{psw}
\ee
where $\phi _{sw}$ is the superweak phase \cite{pdg96}.
Limits on $(g-\overline{g})_J$
for interactions of any sufficiently large range 
may be obtained
by searching for possible modulations of $\metapm $ and $\phipm $ due to
changes in the effective potential,
e.g., as the Earth orbits the Sun, which would induce an
annual modulation in $\phipm $ and $\metapm $. 

From the expressions for $\delta m_{eff}$, quoted above, we can 
calculate the variations of $\phipm $ and $\metapm $
with the variations $\Delta U$ of the astrophysical 
potentials, as determined
from astrophysical measurements \cite{Dekel}.  
One thus obtains the following explicit expressions for
$(g-\overline{g})_J$
\be
|{g-\overline{g}}|_J =
2\sqrt{2}c^2\frac{\deltam }{\Mkao}
\left( \frac{\Delta(\metapm^2)}{\Delta (U^2)}\right) ^{\frac{1}{2}}\xi_{J}
\e ^{r/r_J} 
\label{ggeta2}
\ee
and
\be 
(g-\overline{g})_J =
2\sqrt{2}c^2\frac{\deltam}{\Mkao}\frac{|\epsilon|}{(1+\tan^2\phi_{sw})}
\frac{\Delta\tan\phipm}{\Delta U}\xi_{J} \e ^{r/r_J}~,
\label{ggphi2}
\ee
where $\Delta(\metapm^2 )$ and $\Delta\tan\phipm $ are 
the variations in the observable parameters associated with a variation
$\Delta U$, in the gravitational potential, and
$\xi_{J} =  1, \gamma^{-1}$ and $[\gamma^2(1+v^2/c^2)]^{-1}$ are
the energy dependences for interactions with $J=0, 1$ and $2$. 
It is clear that the sign of $(g-\overline{g})_J$ may be determined
from Eq. (\ref{ggphi2}), but this is not the case for a
measurement of $\metapm $, as can be seen from Eq. (\ref{ggeta2}).
On the other hand, the numerical sensitivity of (\ref{ggeta2}) is
somewhat greater, as we shall see later. The limits quoted in the
following refer always to $|g-\overline{g}|_J$.

We note that Eqs. (\ref{ggeta2}) and (\ref{ggphi2}) depend on $\Delta m$,
for which we take the world average value
$\Delta m= (3.491 \pm 0.009) \times 10^{-12}~{\rm MeV}$ \cite{pdg96}.
In evaluating (\ref{ggphi2}) we take
$|\epsilon |= (2.295 \pm 0.083) \times 10^{-3}$, 
and the superweak phase $\phi_{sw} = 43.49^0 \pm 0.08^0$ \cite{pdg96}.
A straightforward calculation shows that 
the gravitational corrections in $\deltam $, $\phi _{sw}$ and 
$|\epsilon |$ are suppressed by higher powers of $|g - {\overline g}|_J$.
Hence their neglect 
is justified by the small upper limits that we find below on
$|g - \bar{g}|_J$ from searches for modulations in $|\eta_{+-}|$
and $\phipm $.
We also note that the present experimental measurements 
of $\epsilon '$ \cite{epsilonprime} permit us to neglect possible 
direct CP violation in the \Pkao (\Pkab ) $\rightarrow
\pi^+\pi^-$ decay. 

\section{Experimental Search for Violations of the Equivalence Principle}

Our primary limits on $|g-{\overline g}|_J$ will be based 
on possible annual, monthly and diurnal
modulations of $\metapm $ and $\phipm$, associated
with potentials generated by the Sun, Moon 
and Earth; we comment also on possible energy-dependent effects
in the observables (\ref{etapm}) and (\ref{psw}), which could
be present for vector and tensor interactions. Finally, we
give limits on $|g-{\overline g}|_J$  based on
the experimental upper limit of the effective mass difference
$|\delta m|$, in association with galactic and extra-galactic 
potentials.

The data used for this analysis comprise the full data set of 70 million
$\Pkao (\Pkab )\rightarrow \pppm $ decays collected in the  
CPLEAR experiment in the years 1993, 1994 and 1995. The CPLEAR experiment 
is described in detail elsewhere \cite{detector}. 
The CP-violation parameters $\metapm $ and $\phipm $ were determined from the
asymmetry $A_{+-}(\tau)$ formed from the measured numbers of \Pkao\ and
\Pkab\ decaying to $\pppm $, $N(\tau )$ and $\Nb (\tau )$, as functions of
the decay time $\tau $~\cite{cplearcptqmm, pipipaper}:
\begin{eqnarray}
  A_{+-}(\tau ) & = & \frac{\Nb (\tau ) - \alpha N(\tau )}
  {\Nb (\tau ) + \alpha N(\tau )} \nonumber \\
  & = & -2 \frac{\metapm \e ^{\frac{1}{2}(\gs - \gl )\tau}
  \cos (\deltam \tau - \phipm )}{1+ \metapm^2 \e ^{(\gs -\gl )\tau }}~,
\end{eqnarray}
where $\deltam $ is the $\Pkal -\Pkas $ mass difference
and $\gl (\gs )$ the \Pkal (\Pkas ) decay width,
and the normalization factor $\alpha $ is defined in Ref. \cite{pipipaper}.

\begin{figure}[t]                          
  \begin{center}
    \epsfig{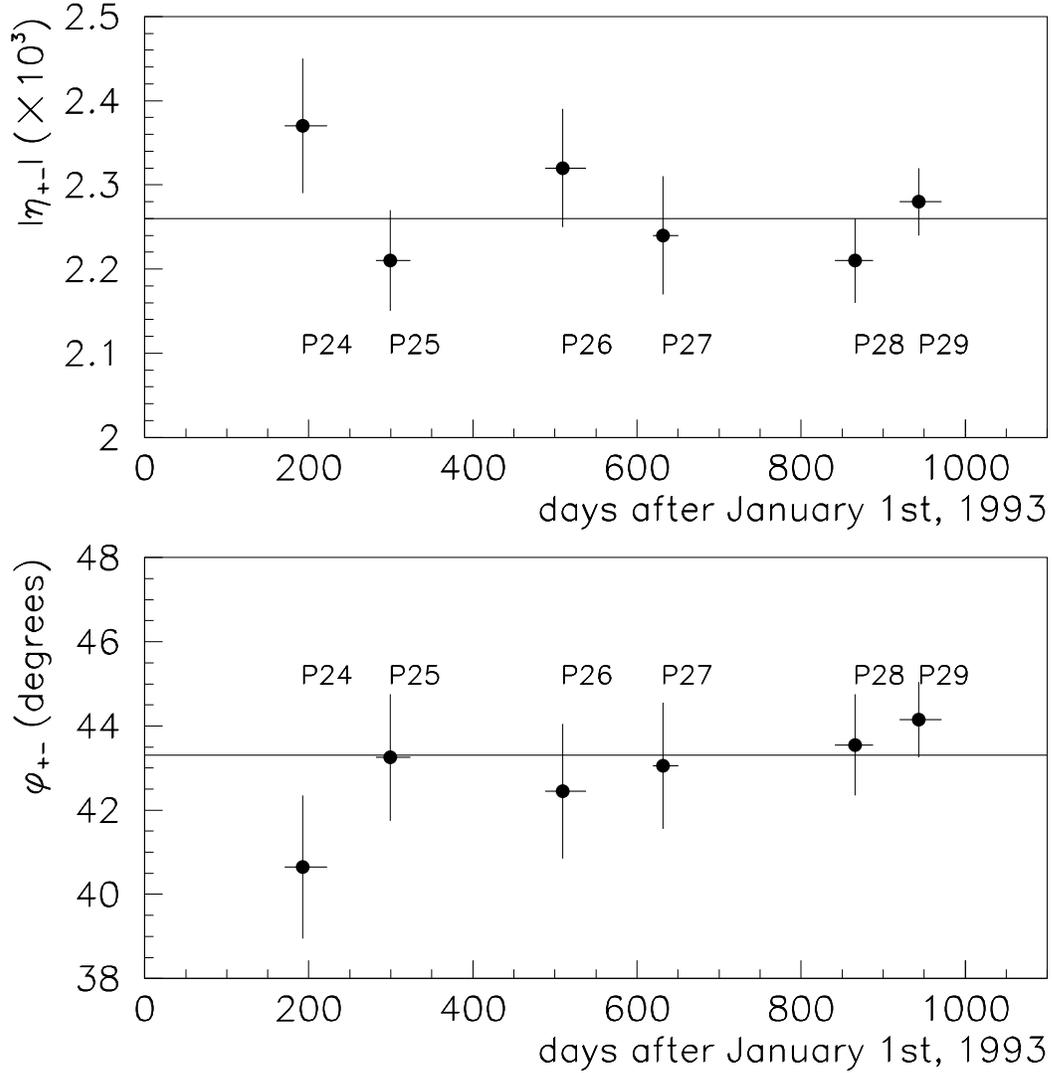}    
    \caption[]{\it The values of $\metapm $ and $\phipm $ 
      measured during the calendar years 1993, 1994 and 1995,
      demonstrating the long-term stability of the detector during different
      running periods (P24-P29). The horizontal error bars correspond to the
      duration of each period and the centre is its mean weighted by the
      number of events collected per day. 
      For each observable, the solid line is the
      result of the fit with a constant and corresponds to
      our average value. The $\chi ^2$/ndf values
      are 4.7/5 and 3.7/5, respectively, for $\metapm $ and
      $\phipm $.}
    \label{fig:stability}
  \end{center}
\end{figure}

We have 
verified that the performance of the detector has been uniform for
the duration of the data-taking periods.  This stability was 
monitored by measuring the mass and width of both  charged and neutral 
$\mathrm{K^*}$s, decaying into a kaon and a pion, for each of the years 1993, 
1994 and 1995 separately. These quantities were found to be stable to 
the level of $10^{-3}$. Moreover we display in Fig.~\ref{fig:stability}
the values of $\metapm $ and $\phipm $ found using data from different
running periods during these years. We note that many effects that
might cause time variations, faking a correlation in the 
performance of the CPLEAR detector, cancel between particles and
antiparticles, which is one of the design features of CPLEAR~\cite{cplear}.

The data are then split into subsamples corresponding to different values of 
the gravitational potential of the Sun, Moon and Earth.
Although the systematic errors on  $\metapm $ and
$\phipm $ are the same for all the subsamples, for the 
purpose of the present study a number of checks were performed
on sources of systematic errors that might induce false correlations
with external variables. These include possible variations in the
size and energy scales of the detector due to temperature or other
effects that could mimic annual or diurnal modulations in the detector
response. We conclude that all the identified systematic errors
are negligible compared with the statistical errors in each of the searches
itemized below.

In view of these searches,
a look-up table was generated from the CPLEAR logbook data, which 
associate run number, date and time with each data-tape written.
The run number of every event is stored with it. The date and time of
each run was used to produce the look-up table, which provides the
gravitational potential for each run number, corrected for the
position of the experiment at that time and the precession and
nutation of the Earth's axis.

Values of  $\metapm $ and
$\phipm $ were  calculated for different subsamples. Limits
on $|g-\overline{g}|_J$ were determined from the slopes of linear fits
with Eqs. (\ref{ggeta2}) and (\ref{ggphi2}) to the data points. All
limits are given at a $95\%$ CL.  Note that in (\ref{ggeta2})
only a non--negative value of $|{g-\overline{g}}|_J$ is allowed.
When we get a negative central value from a fit, we use the tables
in~\cite{feldman} to translate it to a $95\%$ CL limit.

\begin{figure}                           
    \begin{center}
      \epsfig{file=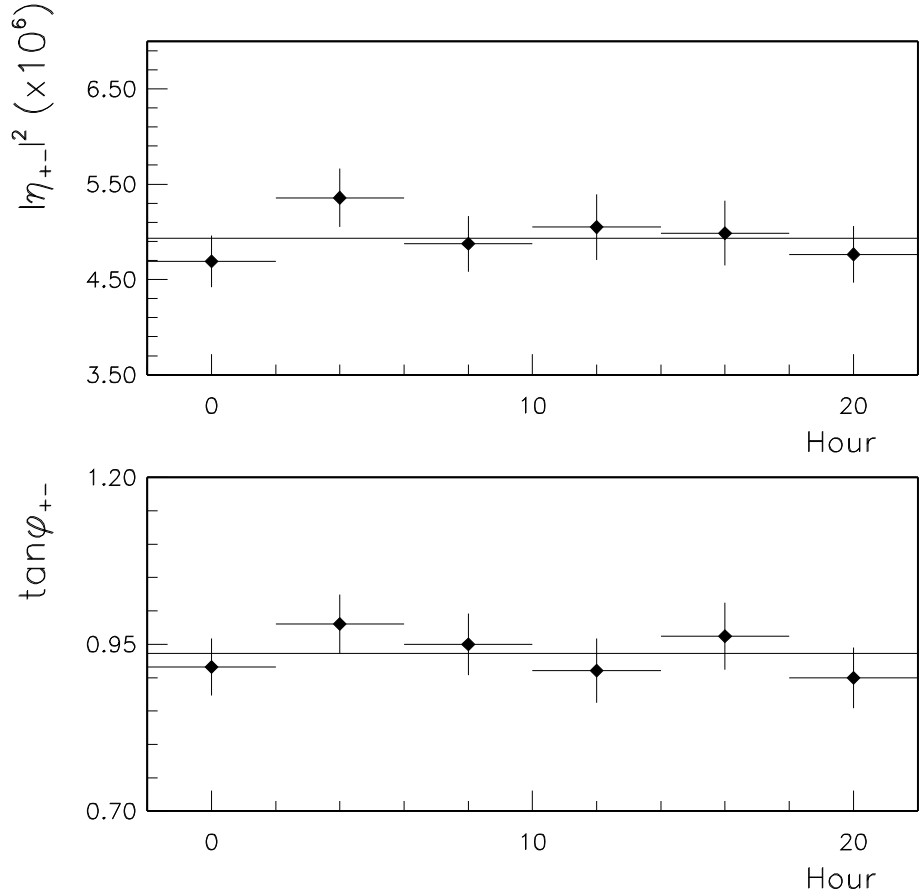,height=14cm}
      \caption[]{\it Measurements of $\metapm ^2$ and $\tan\phipm $
        as functions of the time of day. For each observable, the
        solid line is the
        result of the fit with a constant. The $\chi ^2$/ndf values
        are 3.2/5 and 2.6/5, respectively, for $\metapm ^2$ and
        $\tan\phipm $. }
      \label{fig:ggday}
    \end{center}
  \end{figure}

\begin{figure}[t]                       
    \begin{center}
      \epsfig{file=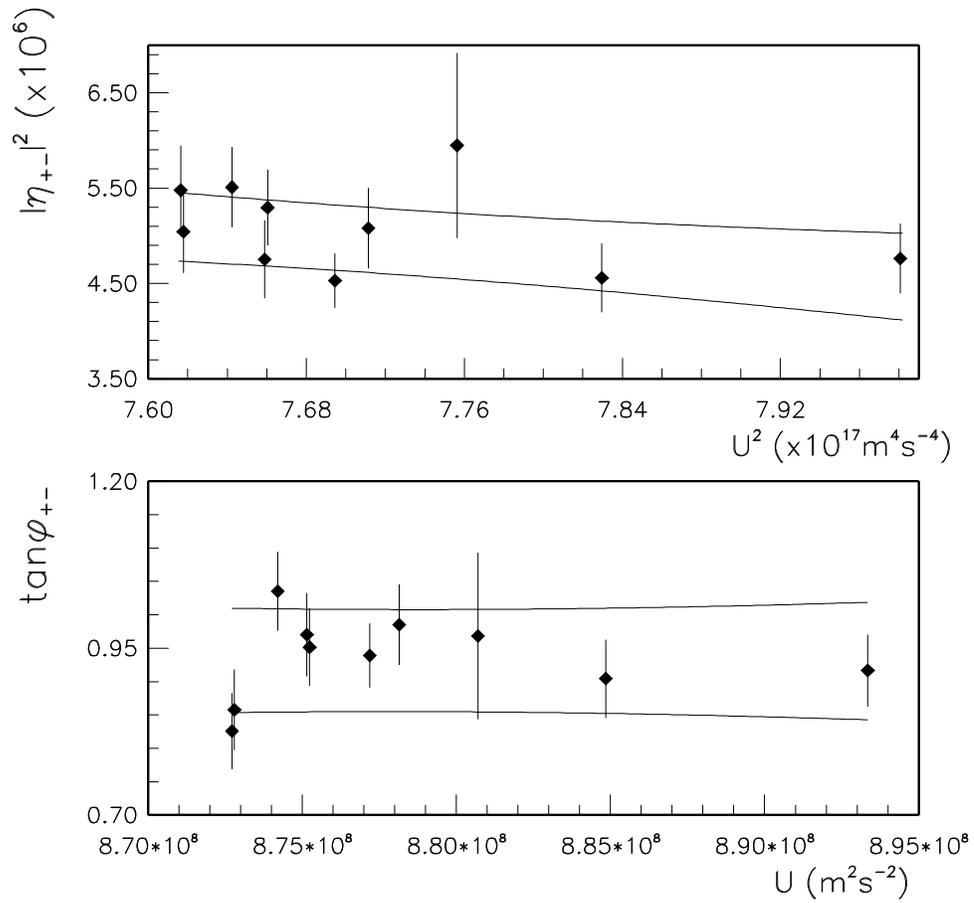,height=14cm}
      \caption[]{\it Measurements of $\metapm ^2 $ and $\tan\phipm $
        as functions of the gravitational potential of the Sun.
        The lines represent the $\pm\, 1\,\sigma$ limits of the region
        around the central values given by the fit. Note that in
        the top plot $(\metapm^2)$ the physical value of the slope cannot
        be negative, as can be seen from Eq. (\ref{ggeta2}).}
      \label{fig:ggsun}
    \end{center}
  \end{figure}

\begin{figure}[t]                        
    \begin{center}
      \epsfig{file=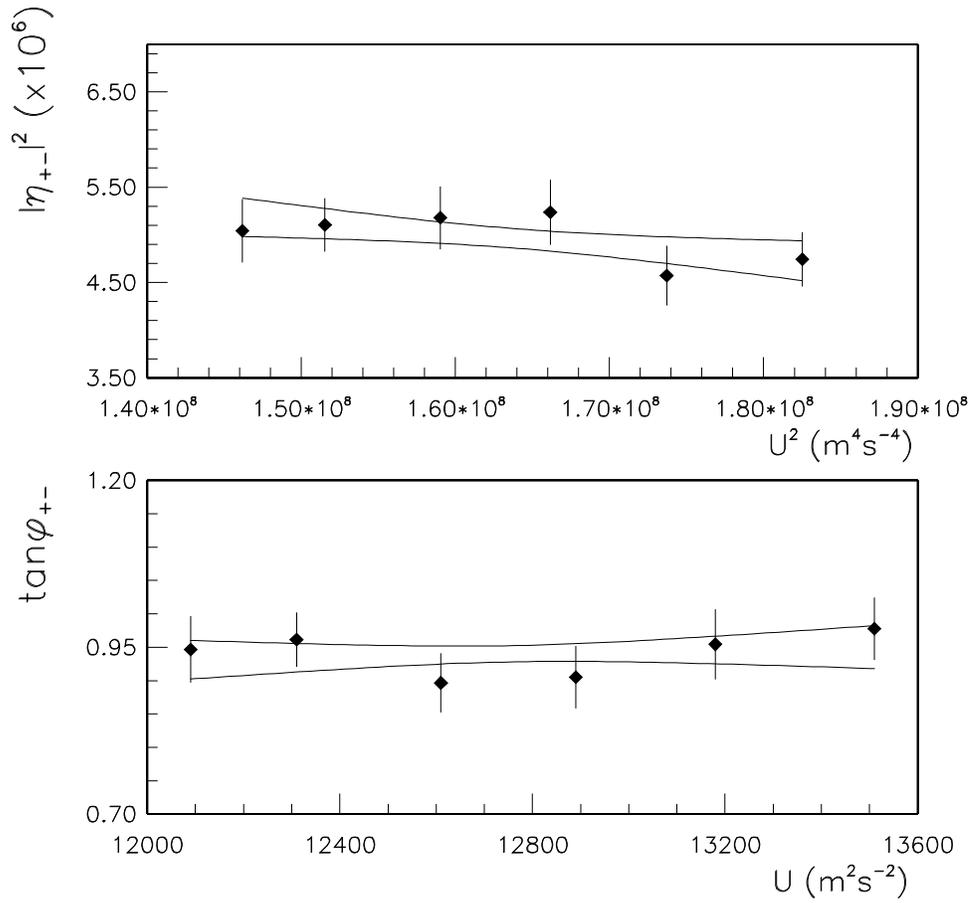,height=14cm}
      \caption[]{\it Measurements of $\metapm ^2 $ and $\tan\phipm $
        as functions of
        the gravitational potential of the Moon and the corresponding
        $\pm 1 \sigma$ regions around the central values given by the fit.}
     \label{fig:lunar}
    \end{center}
  \end{figure}

We have searched for the following possible effects, first
giving limits for a  spin-0 interaction of infinite range, later
for spin-1 and spin-2 interactions and  finally presenting
a compilation in which finite-range effects are taken into account:  
\begin{itemize} 
\item{} A possible diurnal effect:

  As a systematic check, the data were split into six separate samples 
  according to the time of day in order to search for any variation 
  in $\metapm $ and  $\phipm $. The measured values of
  $\metapm $ and $\phipm $ are plotted versus the time of day
  in four-hour bins in Fig.~\ref{fig:ggday}.
  Any astrophysical effect here is expected to be
  much smaller than in the other cases, and indeed none is found.
  We do not use these data to extract upper limits on
  $|g-\overline{g}|_J$.
\item{} Annual modulation due to the Sun's gravitational potential:

  The CPLEAR data were analyzed with respect to the time of year,
  in order to investigate any possible correlation with the
  variations $\Delta U$ in the gravitational potential 
  of the Sun, due to the eccentricity of $0.0167$ in the Earth's orbit.
  The total of the data was split in ten samples  of roughly
  7 million events each. 
  The quantities $\metapm $ and $\phipm $ were calculated for each
  data sample and  the mean time for the sample was obtained as
  a weighted mean. The Sun's distance at this time was used to
  find the corresponding mean gravitational potential.
  The results for $\metapm $ and $\phipm $ are shown in
  Fig.~\ref{fig:ggsun} as functions of the Sun's gravitational potential. 
  Within statistical errors,
  no significant correlation is seen between $\metapm $ or
  $\phipm $ and the potential. The slopes from the 
  fits and their errors are used with Eq. (\ref{ggeta2}) for
  $\metapm $ and Eq. (\ref{ggphi2}) for $\phipm $ to
  calculate limits on $|g-\overline{g}|_0$ of $6.5\times 10^{-9}$
  and $1.2\times 10^{-8}$, respectively. The fit results
  (one-$\sigma$ bands) are also shown in Fig.~\ref{fig:ggsun}.
\item{} Monthly modulation due to the Moon's gravitational potential:

  A search was made for possible monthly variations 
  of $\metapm $ and $\phipm $ with the 
  changing gravitational potential due to the Moon.
  Results are shown in Fig.~\ref{fig:lunar}.
  No significant correlation is observed for 
  either $\metapm $ or $\phipm $, leading to upper limits on
  $|g-\overline{g}|_0$ of $2.0\times 10^{-4}$ and $1.8\times 10^{-4}$
  from $\metapm $ and $\phipm $ , respectively.
\item{} The Earth's gravitational potential:
 
  The data have been split into two subsamples, for neutral kaons travelling
  towards or away from the earth (upwards or downwards).
  No significant variation in $\metapm $ or $\phipm $ is observed, leading 
  to limits on $|g-\overline{g}|_0$ of $6.4\times10^{-5}$ 
  and $3.7\times 10^{-1}$ respectively.  A systematic check was made by 
  an analogous splitting of the data, but 
  where the neutral kaon travelled horizontally, i.e.,
  perpendicular to the Earth's gravitational field. Again, no significant 
  variation in $\metapm $ or $\phipm $ was found.
\item{} Galactic and extra-galactic gravitational potentials:
 
  The time scale of the CPLEAR experiment is clearly too short to
  measure any variation of $\metapm $ or $\phipm $ with the change in
  potential as the Earth orbits the galactic centre. However, we
  can place a limit on $|g-\overline{g}|$ by considering the difference
  between the energies of
  \Pkao\ and \Pkab\ in a common effective galactocentric
  potential associated to a force with
  galactic range~\cite{Dekel}. We take $|\delta m| =
  (-2.6 \pm 2.8)\times 10^{-19}$ GeV,
  as obtained by CPLEAR \cite{cplearcptqmm}. In this way, we find
  an upper limit on $|g-\overline{g}|_0$ of $1.4\times 10^{-12}$
  for a force with range longer than the distance of the Earth from
  the galactic centre, assuming conservatively a mass of $10^{11}$
  solar masses for the galaxy
 \footnote{The corresponding limit based on
  the Sun potential alone is $|g-\overline{g}|_0 < 0.7\times 10^{-10}$.}. 
  
  \vspace{1mm}
  This type of limit may be extended by considering the possible
  effective potential generated by the Virgo cluster (or the Shapley
  supercluster), which is likely to be the most significant extragalactic
  source. In these cases the relevant masses and distances 
  are less well known. However, these have been estimated to be
  $10^{14}$ $(5 \times 10^{16})$ solar masses and $15-20$ ($250$) Mpc
  respectively~\cite{Dekel}, giving upper limits for $|g-\overline{g}|_0$
  of $0.9\times10^{-12}$ ($0.7 \times 10^{-13}$).
  This may be compared to the similar analysis in~\cite{hughes}.

  \vspace{1mm}
  It should be emphasized once more that this method  
  used for the galactic and supercluster cases
  requires the use of an absolute potential, and hence involves an
  extra theoretical assumption. It should be noted that this 
  method cannot be  applied to  the Sun, Earth  and Moon,
  since presumably the galactic and/or extragalactic potentials dominate
  at the Earth surface, if the range is large. The method of placing
  bounds using  $\phipm $
  and $\metapm $ modulations avoids the use of absolute potentials 
  altogether, and therefore the bounds so obained are the best available
  from a model-independent point of view. 
\end{itemize}
\vspace*{3mm}
In all the cases considered, the values obtained for $\metapm $
and $\phipm $ do not show a significant dependence on
the potential, within 1.5 standard deviations. The above limits
on $|g-\overline{g}|$ refer to a spin-0 interaction. 
Table~\ref{tab:summary} 
summarizes the best limits on $|g-\overline{g}|$ 
obtained from $\metapm $  and $\phipm $ for tensor,
vector and scalar interactions. 
We note that the limits from $\metapm $ are usually 
better than those from the $\phipm $ data.
The limits for galactic
and extragalactic gravitational potentials, obtained from $\delta m$,
are also reported in Table~\ref{tab:summary}.    
The limits given in Table~\ref{tab:summary} were all calculated
for forces of range much longer than the corresponding
astrophysical distance scale,
so that the exponential factors ${\rm exp}(r/r_J) \simeq 1$.
 
It has been suggested previously~\cite{Bell} 
that all the CP violation observed in the neutral-kaon
mass matrix might be due to the interaction with an
astrophysical source. Our results, on the absence
of modulations correlated, e.g., with the Earth-Sun distance
do not allow us to reject this hypothesis, although they
can be used to constrain the allowed couplings of conjectural
particles as functions of their masses \cite{Nieto}.

\begin{table}
  \begin{center}
    \begin{tabular}{||l|r|r|r||} \hline
      Source & Spin 0 & Spin 1& Spin 2 \\ \hline \hline 
      Earth  & $6.4\times10^{-5}$  & $4.1\times10^{-5}$  & $1.7\times10^{-5}$
      \\
      Moon   & $1.8\times10^{-4}$  & $7.4\times10^{-5}$  & $4.8\times10^{-5}$
      \\
      Sun    & $6.5\times10^{-9}$  & $4.3\times10^{-9}$  & $1.8\times10^{-9}$
      \\ \hline \hline
      Galaxy & $1.4\times10^{-12}$ & $9.1\times10^{-13}$ & $3.8\times10^{-13}$
      \\
      Supercluster & $7.0\times10^{-14}$ & $4.6\times10^{-14}$
      &$1.9\times10^{-14}$ \\ 
      \hline
    \end{tabular}
    \caption[]{\it Summary of limits on $|g-\overline{g}|$
     for spin 0, 1 and 2 interactions.}
    \label{tab:summary}
  \end{center}
\end{table}

In the cases of spin-1 and spin-2 interactions, limits can in principle 
also be obtained from studies of the energy dependence of parameters of
the \Pkao -- \Pkab\ system~\footnote{These can also be used to 
constrain the phase difference parameters in the generalized
interaction formalism of~\cite{Hambye} for violations of the
Equivalence Principle.}.
We note that, in view of the small range of $\gamma$ involved 
in the CPLEAR experiment ($\gamma=1.54$ at the average kaon momentum)
and the stringent
upper limits on $|g-\overline{g}|$ obtained above, CPLEAR is not
sufficiently sensitive to constrain
significantly interactions with the energy 
dependences given in (\ref{diff}) and (\ref{diffvect}).
However, we point out that
a similar analysis of the combined data taken by CPLEAR and
experiments E731, E773 at FNAL~\cite{energ} would have considerably
greater sensitivity to spin-1 and particularly spin-2 interactions,
since the overall range of $\gamma$ would extend up to the order of $10^6$.

\begin{figure}[t]                         
  \begin{center}
    \epsfig{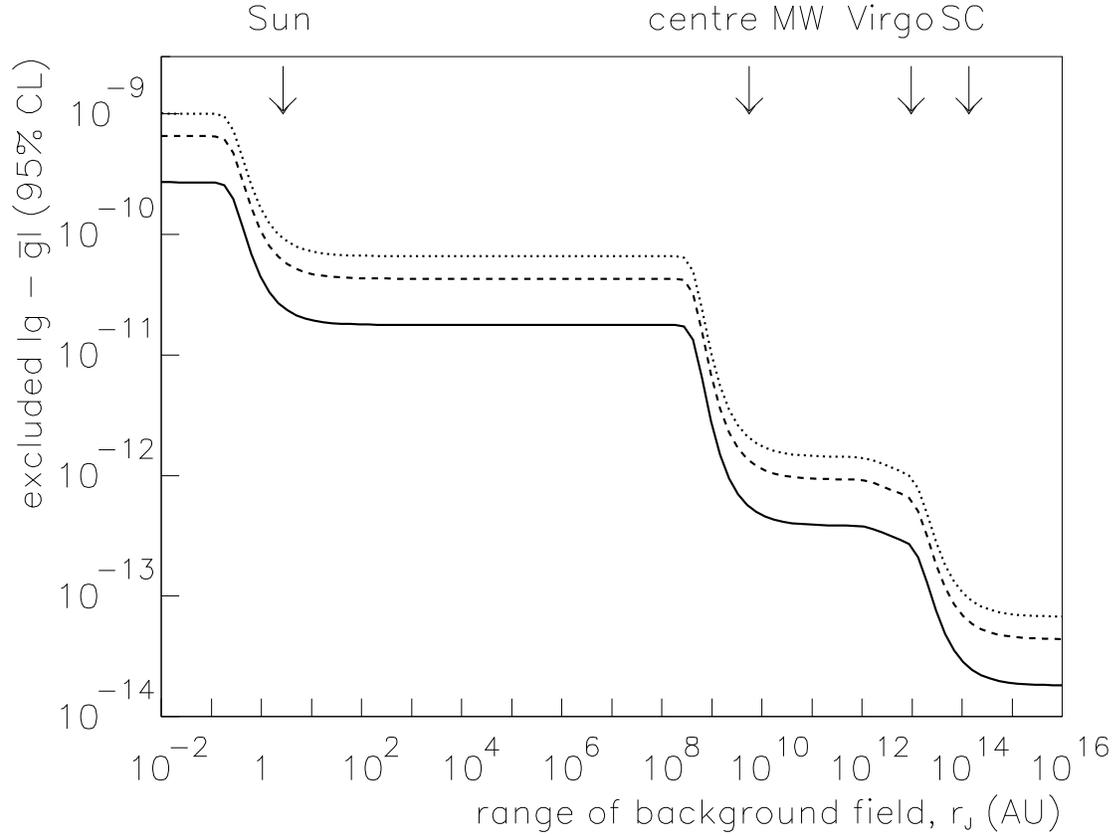}
    \caption[]{\it Limits on $|g-\overline{g}|_J$ 
      arising from the measured  \Pkao -- \Pkab\ mass difference
      \cite{cplearcptqmm} as a function of 
      the effective interaction range. Labels along the top
      denote the distances to several astronomical bodies (Milky Way:
      MW, Shapley supercluster: SC) measured in Astronomical units (AU).
      The curves are upper limits shown separately for tensor (solid
      line), vector (dashed line) and scalar (dotted line) interactions.}
    \label{fig:parabs}
  \end{center}
\end{figure}

Finally, we use the limit on the \Pkao -- \Pkab\ mass difference
determined by CPLEAR \cite{cplearcptqmm} and  the full functional
form of Eqs. (\ref{diff}) -- (\ref{diffscal}) to compile
the limits on $|g-\overline{g}|_J$ for a spin-$J$ interaction
as a function of the finite range $r_J$. These limits are shown
in Fig.~\ref{fig:parabs}. It should be emphasized that
the more stringent limits at large range have intrinsically larger
uncertainties, associated with uncertainties in modelling large-scale
structures. 
\section{Conclusion} 

We have found no evidence for any variation of $\metapm $ or $\phipm $ 
associated with possible effective potentials generated by the Earth,
Moon or Sun. We have used our data to establish stringent upper limits
on possible effective spin-0, -1 or -2 interactions that might induce
apparent deviations from the Principle of Equivalence, which we have
given as functions of their possible ranges.

\section{Acknowledgements}

We would like to thank the CERN LEAR staff for their support and
co-operation, as well as the technical and engineering staff of
our institutes. 
We would also like to thank M. Currie and J. Lockley at Starlink (the 
computing facility for UK astronomers) for their help with Starlink's
software.
We thank K.~Zioutas for his interest, and also
thank M. Davis, A. Dekel and A. Eldar for their advice. 
This work was supported by the following agencies: 
the French CNRS/Institut National de
Physique Nucl\' eaire et de Physique des Particules (IN2P3),
the French Commissariat \` a l'Energie Atomique (CEA),
the Greek General Secretariat of Research and Technology,
the Netherlands Foundation for Fundamental Research on Matter (FOM),
the Portuguese JNICT,
the Ministry of Science and Technology of the Republic of Slovenia,
the Swedish Natural Science Research Council,
the Swiss National Science Foundation,
the UK Particle Physics and Astronomy Research Council (PPARC),
the US National Science Foundation (NSF) and 
the US Department of Energy (DOE).

\begin{thebibliography}{99}
\bibitem{cptkaon}
      J.S. Bell, Proc. Royal Soc. {\bf A231} (1955), 479;\\
      W. Pauli, in {\it Niels Bohr and the Development of 
      Physics}, eds. W. Pauli, L. Rosenfeld and V. Weisskopf (Mc Graw Hill,
      New York, 1955);\\
      G. L\"uders, Ann. Phys. (N.Y.) {\bf 2} (1957), 1;\\
      R. Jost, Helv. Phys. Acta {\bf 30} (1957), 409.  
\bibitem{cptconv} N. Tanner and R. Dalitz, Ann. Phys. (N.Y.)
      {\bf 171} (1986), 463;\\
      C.D. Buchanan {\it et al.}, Phys. Rev. {\bf D45} (1992), 4088;\\
      C.O. Dib and R.D. Peccei, Phys. Rev. {\bf D46} (1992), 2265. 
\bibitem{pdg96} R.M. Barnett {\it et al.}, Particle Data Group,
      Phys. Rev. {\bf D54} (1996), 412.
\bibitem{cplearcptqmm} CPLEAR Collaboration,
      {\it Results on CP, T, CPT symmetries with tagged
      \Pkao\ and \Pkab\ by CPLEAR}, Proc. ICHEP'98, Vancouver,
      1998, and papers to be published.
\bibitem{ehns} J. Ellis, J. Hagelin, D.V. Nanopoulos and M. Srednicki,
      Nucl. Phys. {\bf B241} (1984), 381.  
\bibitem{emn} J. Ellis, N.E. Mavromatos and D.V. Nanopoulos, 
      Phys. Lett. {\bf B293} (1992), 142;
      Int. J. Mod. Phys. {\bf A11} (1996), 1489;\\
      J. Ellis, J. Lopez, N.E. Mavromatos and D.V. Nanopoulos,
      Phys. Rev. {\bf D53} (1996), 3846;\\
      P. Huet and M.E. Peskin, Nucl. Phys. {\bf B434} (1995), 3.
\bibitem{cptcplear}R. Adler {\it et al.}, CPLEAR Collaboration, and  
      J. Ellis, J. Lopez, N.E. Mavromatos and D.V. Nanopoulos,
      Phys. Lett. {\bf B364} (1995), 239. 
\bibitem{cptgrav} R. Wald, Phys. Rev. {\bf D21} (1980), 2742;\\
      D. Page, Gen. Rel. Grav. {\bf 14} (1982), 299. 
\bibitem{graviphoton} J. Scherk, Phys. Lett. {\bf B88} (1979), 265;
      see also:\\ S. Belucci and V. Faraoni,
      Phys. Rev. {\bf D49} (1994), 2922;
      Phys. Lett. {\bf B377} (1996), 55, and  references therein. 
\bibitem{equiv} M.L. Good, Phys. Rev. {\bf 121} (1961), 311.
\bibitem{su94} Y. Su {\it et al.}, Phys. Rev. {\bf D50} (1994), 3614.
\bibitem{will96}J.G. Williams {\it et al.},
      Phys. Rev. {\bf D53} (1996), 6730.
\bibitem{ves80} R.F.C. Vessot {\it et al.},
      Phys. Rev. Lett. {\bf 45} (1980), 2081.
\bibitem{chup89}T.E. Chupp {\it et al.},
      Phys. Rev. Lett. {\bf 63} (1989), 1541;\\
      S.K. Lamoreaux {\it et al.},
      Phys. Rev. Lett. {\bf 57} (1986), 3125;\\
      J.D. Prestage {\it et al.},
      Phys. Rev. Lett. {\bf 54} (1985), 2387.
\bibitem{pres95} J.D. Prestage {\it et al.},
      Phys. Rev. Lett. {\bf 74} (1995), 3511.
\bibitem{gab90} G. Gabrielse {\it et al.},
      Phys. Rev. Lett. {\bf 65} (1990), 1317.
\bibitem{hug91} R.J. Hughes and M.H. Holzscheiter,
      Phys. Rev. Lett. {\bf 66} (1991), 854. 
\bibitem{Bell} J.S. Bell, in {\it Fundamental Symmetries},
      eds. P. Bloch, P. Pavlopoulos and R. Klapisch
      (Plenum, New York, 1987); for an alternative formulation, see\\
      G. Chardin and J.-M. Rax, Phys. Lett. {\bf B282} (1992), 256;\\
      G. Chardin, Nucl. Phys. {\bf A558} (1993), 477c;
      Hyp. Int. {\bf 109} (1997), 83.
\bibitem{hughes} R.J. Hughes, Phys. Rev. {\bf D46} (1992), R2283.
\bibitem{Dekel} M. Davis, M.A. Strauss and A. Yahil, Ap. J. {\bf 372}
      (1991), 394;\\
      M. Davis, A. Dekel and A. Eldar, private communication.
\bibitem{epsilonprime} P. Buchholz {\it et al.},
      Phys. Lett. {\bf B317} (1993), 233;\\
      A.P. Barker {\it et al.},
      Phys. Rev. Lett. {\bf 70} (1993), 1203.
\bibitem{detector} R. Adler {\it et al.}, CPLEAR Collaboration,
      Nucl. Instr. \& Meth. {\bf A379} (1996), 76.
\bibitem{pipipaper}R. Adler {\it et al.}, CPLEAR Collaboration,
      Phys. Lett. {\bf B363} (1995), 243. 
\bibitem{cplear} E. Gabathuler and P. Pavlopoulos, in 
      {\it Strong and Weak CP Violation at LEAR},
      Proc. Workshop on Physics at LEAR with Low Energy
      Cooled Antiprotons, Erice, 1982, eds. U. Gastaldi and R. Klapisch
      (Plenum, New York, 1984), p.747.
\bibitem{Nieto} M.N. Nieto and T. Goldman,
      Phys. Rep. {\bf 5} (1991), 221;\\
      R.J. Hughes, T. Goldman and M.N. Nieto,
      in {\it Fundamental Symmetries},
      eds. P. Bloch, P. Pavlopoulos and R. Klapisch
      (Plenum, New York, 1987).
\bibitem{feldman} G.J. Feldman and R.D. Cousins,
      Phys. Rev. {\bf D57} (1998), 3873.
\bibitem{Hambye} T. Hambye, R.B. Mann and U. Sarkar, 
      hep-ph/9608423; Phys. Lett. {\bf B421} (1998),105; 
      Phys. Rev. {\bf 58} (1998) 025003.
\bibitem{energ} L.K. Gibbons {\it et al.},
      Phys. Rev. Lett. {\bf 70} (1993), 1199;\\
      B. Schwingenheuer {\it et al.},
      Phys. Rev. Lett. {\bf 74} (1995), 4376. 
  
\end{thebibliography}
\end{document}